\documentclass[aps,prl,twocolumn,groupedaddress]{revtex4}
\usepackage{color}
\usepackage{epsf}
\usepackage{epsfig}
\usepackage{graphicx}
\usepackage{latexsym}
\usepackage{bm}
\usepackage{amssymb}
\usepackage[
   colorlinks,
   linkcolor=black,
   urlcolor=blue,
   citecolor=blue
   ]{hyperref}

\newcommand{\Fref}[1] {Figure \ref{#1}}

\begin{document}

\draft

\preprint{to be submitted to PRL}

\title {Ejection of quasi-free electron pairs from the helium atom ground
state by single photon absorption}

\author{M.~S.~Sch\"{o}ffler$^{1}$}
\email{schoeffler@atom.uni-frankfurt.de}
\author{C.~Stuck$^{1,2}$}
\author{M.~Waitz$^2$}
\author{F.~Trinter$^2$}
\author{T.~Jahnke$^2$}
\author{U.~Lenz$^2$}
\author{M.~Jones$^3$}
\author{A.~Belkacem$^{1}$}
\author{A.~Landers$^{3}$}
\author{M.~S.~Pindzola$^{3}$}
\author{C.~L.~Cocke$^{4}$}
\author{J.~Colgan$^{5}$}
\author{A.~Kheifets$^{6}$}
\author{I.~Bray$^{7}$}
\author{H.~Schmidt-B\"{o}cking$^2$}
\author{R.~D\"{o}rner$^2$}
\author{Th.~Weber$^1$}

\affiliation{$^1$ Lawrence Berkeley National Laboratory, Berkeley,
CA 94720, USA}

\affiliation{$^2$ Institut f\"ur Kernphysik, University Frankfurt,
Max-von-Laue-Str. 1, 60438 Frankfurt, Germany}

\affiliation{$^3$ Department of Physics, Auburn University, Auburn,
AL 36849, USA}

\affiliation{$^4$ Department of Physics, Kansas State University,
Manhattan, KS 66506, USA}

\affiliation{$^5$ Theoretical Division, Los Alamos National
Laboratory, Los Alamos, NM 87545, USA}

\affiliation{$^6$ Research School of Physical Sciences and
Engineering,
Australian National University, Canberra, ACT 0200, Australia}

\affiliation{$^7$ ARC Centre for Antimatter-Matter Studies, Curtin
University, WA 6845 Perth, Australia }

\vskip 5mm

\date{\today}

\vskip 5mm

\begin{abstract}

We investigate single photon double ionization (PDI) of helium at
photon energies of 440 and 800~eV. We observe doubly charged
ions with close to zero momentum corresponding to electrons emitted
back-to-back with equal energy. These slow ions are the unique
fingerprint of an elusive quasi-free PDI mechanism predicted by Amusia
{\em et al.} nearly four decades years ago [J. Phys. B \textbf{8,}
1248, (1975)] . It results from the non-dipole part of the
electromagnetic interaction. Our experimental data are in excellent
agreement with calculations performed using the convergent close
coupling and time dependent close coupling methods.

\end{abstract}


\maketitle


\section{Introduction \label{introduction}}

The non-relativistic Hamiltonian of electromagnetic interaction is a
one-body operator and thus it couples one photon to just one
electron. Simultaneous emission of two electrons following absorption
of a single photon is only possible due to electron-electron
correlation.  Single photon double ionization (PDI) of helium became
the prototype process to investigate such correlations. The two
specific correlation mechanisms, the shake-off (SO) and electron
knock-off (KO) processes (the latter is also known as Two-Step-1 or
TS1 \cite{McGuire97}), are well established today to facilitate the
ejection of two electrons by a single photon. In both cases, the
photon couples primarily to the dipole formed by one electron and the
nucleus. The two electrons interact with each other, either before
(SO) or after (TS1) the photo-absorption takes place. At photon
energies far from the threshold, the electron emission
patterns characteristic for each of the mechanisms can be separated in fully differential cross sections (FDCS)
\cite{Knapp2002prl}. The energy sharing between the two emitted
electrons exhibits a U-shape
\cite{Schneider2002prl,Schneider2003pra}, which becomes deeper and
deeper with increasing photon energies \cite{Wehlitz1991prl,Proulx1995pra}. The TS1 probability decreases
with the increasing photon energy while the SO probability increases
and finally saturates at the (non-relativistic) shake-off limit
\cite{Aberg1967pr, Spielberger1995prl, Burgdoerfer1998pra, Kheifets2001jpb}. A common
fingerprint of both mechanisms is a large momentum of the doubly
charged ion and an almost dipolar angular distribution of the ion
momentum with respect to the linear photon polarization axis. These two
characteristic features are signatures of the initial coupling of the photon
to the dipole formed by the primary electron and the nucleus \cite{Knapp2002prl, Spielberger1995prl, Knapp2005jpb1, Doerner1996prl, Braeuning1997jpb}.

Nearly four decades ago, \citet{Amusia1975jpb} predicted a third, so
called quasi-free mechanism (QFM) of PDI. Its characteristic
fingerprint is the ejection of two electrons back-to-back with similar
energy while the nucleus is only a spectator \cite{Ludlow2009jpb}
remaining almost at rest \cite{Amusia1975jpb, Amusia2003jpb, Amusia2012jetp}. They
argued that this mechanism ejects electrons mainly from the part of
the initial state wave function at the electron-electron cusp,
i.e. where both electrons are spatially close together. It is a
contribution to the quadrupole part of PDI since emitting electrons
back-to-back with equal energies is forbidden by the kinematic
selection rule in a dipole transition \cite{Maulbetsch1995jpb}. This
region of momentum space is, however, allowed in a quadrupole
transition.

The QFM transition amplitude is extremely small, which is why it could
not be verified experimentally so far. For instance, for a photon
energy of 800 eV and the hydrogen Bohr radius, the non-dipole
transition amounts to 1~\% of the total PDI cross section only. From
this quadrupole part, the QFM is only a small part; it can be
estimated to 0.1~\% of the total PDI cross section.

PDI of He in the low photon energy regime, where quadrupole
transitions are negligible, has been investigated experimentally and
theoretically in great detail in the past
\cite{Briggs2000jpb,Malegat2004ps,Avaldi2005jpb}
Experimental FDCS are available up to 530~eV \cite{Knapp2002prl,
Knapp2002jpb, Knapp2005jpb1, Knapp2005jpb2, Knapp2005jpb3}. However,
very little is known about PDI at higher photon energies. The coupling
of higher orders of angular momentum of the incoming light, such as
the electric quadrupole term, to the two electrons in the atom has
only been addressed theoretically \cite{Mikhailov2003pla, Mikhailov2004pra, Istomin2004prl,
Istomin2005pra, Popov2012arxiv, Amusia2011arxiv}. So far, no
experiments have studied non-dipole effects in PDI due to extremely
small cross sections. As the QFM has an even smaller cross section and
requires, in addition, the coincident detection of two electrons
emitted back-to-back with equal energy, it escaped experimental
observation until now.

In the present Letter, we report on the observation of low momentum
doubly charged ions for PDI of helium at 800~eV and thus present
direct support for the QFM mechanism. This is also consistent with the accompanying {\em ab initio}
non-perturbative calculations, performed using the convergent close
coupling (CCC) and the time dependent close coupling (TDCC) methods,
which reproduce our experimental findings.

\begin{figure}[htbp!]
 \begin{center} \epsfig{file=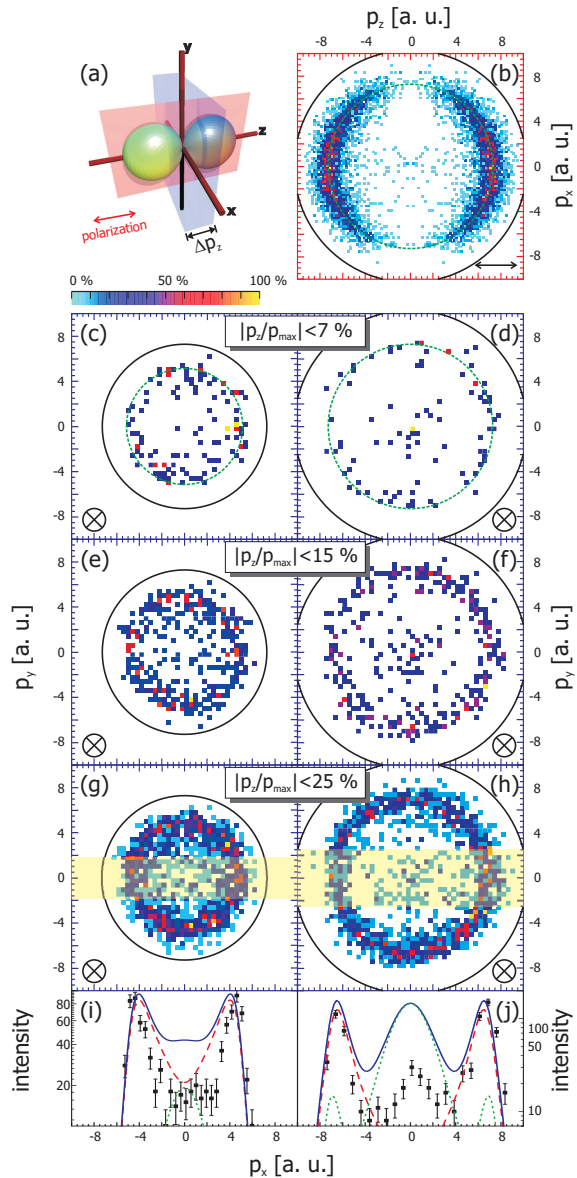,width=7.5cm}
 \caption{(a) Sketch of the dipole structure, different planes and coordinates: light propagation x, light polarization z, gas jet direction y. (b) Slice ($|p_y|<0.73 a. u.$) of the ion momentum distribution in the x,z-plane for $\hbar\nu$=800 eV. (c-h) Ion momentum distribution perpendicular to the light polarization vector for $\hbar\nu$=440~eV (left column) and $\hbar\nu$=800~eV (right column), with different cut widths along the momentum component parallel to the polarization vector of the doubly charged ion at a given photon energy ($p_z$) (c,d) $\left| p_z/p_{max}\right|\leq7\%$, (e,f) $\left| p_z/p_{max}\right|\leq15\%$, (g,h) $\left| p_z/p_{max}\right|\leq25\%$. The solid circular line represents the maximum possible momentum ($p_{max}$). The dashed green line in b-d represents ($p_{single}$), the case if one electron receives all momentum. (i,j) show a momentum cut of (g,h) $\left| p_y/p_{max}\right|\leq25\%$ indicated by the yellow bars. The various lines show TDCC calculations for dipole (dashed and red), quadrupole (dotted and green) as well as the coherent sum of dipole and quadrupole (blue and solid).
\label{rpxpy}}
\end{center}
\end{figure}

The experiments were carried out with linear polarized light at 800~eV
during two-bunch mode at beamline~11.0.2.1 of the Advanced Light
Source (ALS) of LBNL. The technical realization of this project is
very challenging. The cross section for quadrupole transitions amounts
to $\rm \simeq2\cdot10^{-25}cm^2$ only. At the same time, a coincidence measurement of the
momentum vectors of at least one electron and the recoiling doubly
charged ion is necessary to cleanly single out the scarce PDI events
(total cross section $\rm \simeq2\cdot10^{-23}cm^2$). The solution to
this problem is the application of the COLd Target Recoil Ion Momentum
Spectroscopy (COLTRIMS) method which is able to detect the 3D-momenta
of the outgoing particles within the $4\pi$ solid angle in coincidence
\cite{Ullrich1997jpb,Doerner2000pr,Ullrich2003rpp}. It is sufficient
to measure one electron and the recoiling ion simultaneously. Momentum conservation is used to calculate the momentum vector of the second electron and the
energy conservation is exploited to eliminate background in the
offline analysis. In brief, the target is prepared in a supersonic gas
jet (30~$\mu$m nozzle and 20~bar driving pressure) along the
$y$-direction and intersected with the photon beam (propagating along
the $x$-axis) in a weak electric field inside the momentum
spectrometer. The field (16~V/cm) is just strong enough to separate
the fragments by their charge and guide them towards two large position and
time sensitive multi-channel plate (MCP) detectors with delay line
readout \cite{Jagutzki2002nima, Jagutzki2002nima2}; 80~mm diameter for
the ions and 120~mm for the electrons. A magnetic field (23~Gauss) in
parallel is used to prevent high energetic electrons from leaving the
spectrometer by forcing these particles to gyrate towards the
detector. With the knowledge of the dimensions of the spectrometer,
the field strengths, the position of impact and time of flight of the
particles, the 3D momentum vector of each particle can be deduced. To increase the ion momentum resolution and to
compensate for the finite interaction volume (see \cite{Schoeffler2011njp} for
general information about the time/space focusing), we employed an
electrostatic lens and a 120~cm long drift region for focusing. This
resulted in an ion momentum resolution of $\simeq$0.15~a.u. in the
light polarization direction ($z$), which is also the
time-of-flight-direction, and $\simeq$0.25~a.u. in the transverse
direction. The axes layout is shown in \Fref{rpxpy}(a). The photon
beam had a small contamination of lower harmonics. Due to an increase of
the PDI cross section with decreasing photon energy, this small
contamination leads to a non negligible amount of low momentum doubly
charged ions. Our coincidence experiment however provides for each
registered event the sum energy of both electrons. This has been used
to identify the contamination of the ion signal by low energy photons
and to unambiguously select events which resulted from the absorption
of one 800~eV photon. This also discriminates against events from compton scattering, which also would produce low momentum ions \cite{Spielberger1995prl}

In \Fref{rpxpy}(b) we present the momentum distribution of the doubly
charged helium ions after PDI with linear polarized light of
800~eV. The solid circular line represents the maximum possible momentum an ion could receive in a double ionization ($\rm p_{max}=2\sqrt{(E_\gamma - 79 eV)/2}$, where 79 eV represents the PDI threshold). This is the case, when both electrons have half the excess energy and are emitted in the same direction. Nevertheless the vast majority of events are located close to the surface of a sphere in the momentum space with a radius of $\rm p_{single}=\sqrt{E_\gamma - 79~eV}$ which corresponds to the
recoil momentum of one electron that takes away all the available
energy. These ions show a close to dipolar angular distribution. This
is the characteristic pattern observed in all previous experiments
\cite{Doerner1996prl,Knapp2002prl,Braeuning1997jpb}. It is created by
both the SO and TS1 process and dominated by the dipole transition by
far. In the center of the sphere, close to the momentum zero, we
expect the events from the QFM. To make those events more visible, we
plot the two momentum components in the $xy$-plane perpendicular to
the polarization direction (\Fref{rpxpy}c-h).  We define the momentum
width out of this plane ($z$-direction) that we are going to select as
$\left|\Delta p_z/p_{\max}\right|$ and show cuts corresponding to 7,
15 and 25~\% of the maximum momentum in \Fref{rpxpy}c-d, e-f and g-h,
respectively. This allows us to exploit the selection rules for the
dipole transitions which forbid the emission of both electrons in the
$xy$ plane, irrespective of the energy sharing \cite{Maulbetsch1995jpb}. The outer ring of the pattern
corresponds to transitions with maximum unequal electron energy
sharing. In the center of this ring we find ions almost at rest, their
relative contribution increases with photon energy; compare the left
column (440~eV) with the right column (800~eV). As outlined above, we
have performed a kinetically complete experiment, i.e. we obtained the
momentum vectors of all particles. This allows for a full control of
the experiment and a highly efficient suppression of all background. We
therefore can be certain that the few events at small momentum are
experimentally significant and definitely not background. A projection from (\Fref{rpxpy}g-h) with a width of 25~\% (indicated by the yellow bars) is shown in (\Fref{rpxpy}i-j). Also here a peak around zero momentum is clearly visible for 800, but not for 440 eV.

\begin{figure}[htbp!]
  \begin{center}
    \epsfig{file=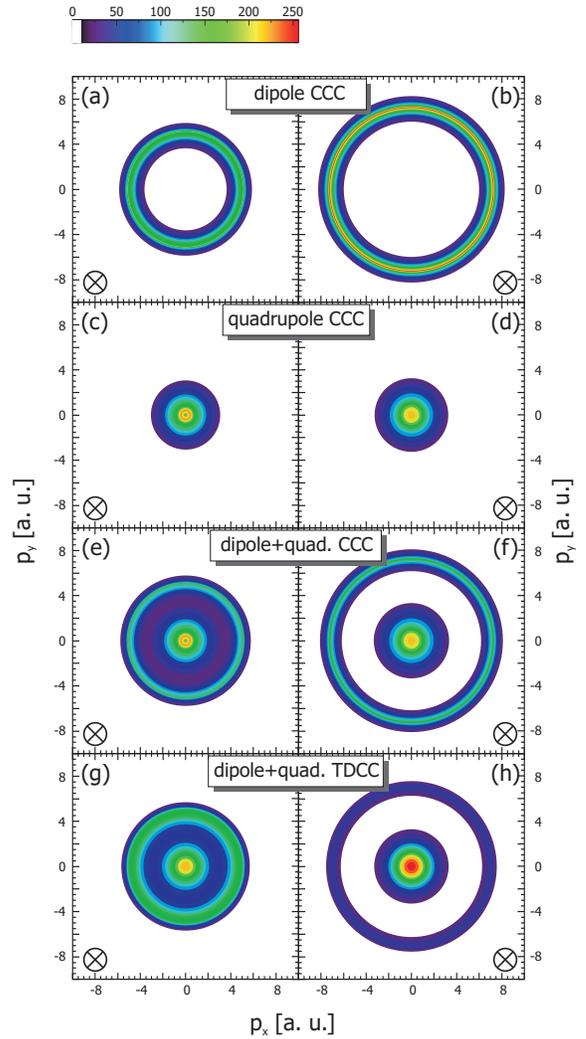,width=7.5cm}
    \caption{Calculations for 440~eV (left column) and 800~eV (right column) photon energy with the same geometry as in the experiment (\Fref{rpxpy}c,d), for a width of $\Delta p_z/p_{\max}\leq\pm15\%$. In panels (a,b) the CCC calculations include only the dipole interaction, while (c,d) display solely the quadrupole terms and (e,f) show the full theory. The panels (g,h) show a full TDCC calculation including dipole and quadrupole terms and in contrast to CCC their interference terms.
    \label{fig2} }
  \end{center}
\end{figure}

The present experimental findings are supported by convergent close
coupling (CCC) calculations \cite{Briggs2000jpb} and time
dependent close coupling (TDCC) theory \cite{Ludlow2009jpb} shown in
\Fref{fig2} for 440~eV (left) and 800~eV (right) photon energies and a
cut in $p_z=\pm15\%$. These calculations contain the dipole part
(\Fref{fig2} a,b) and quadrupole contributions (\Fref{fig2} c,d) as well
as the coherent (TDCC) and incoherent (CCC) sum of the dipole and
quadrupole matrix elements (\Fref{fig2} e,f). Both calculations show
ions with zero momentum for the quadrupole, but not for the dipole
terms. The individual dipole and quadrupole contributions from CCC and TDCC agree well. 
The higher momenta contain contributions from both the dipole
and the quadrupole term. The TDCC is a direct solution of the time
dependent Schr\"odinger question. It is not possible to connect this
to the intuitive picture of ionization mechanisms. As the calculation
is exact up to the quadrupole term, it does, however, include all the
possible PDI mechanisms. The CCC calculations, in turn, can be analyzed in
terms of the Feynman diagrams \cite{Kheifets2001jpb}. The CCC theory does
include the shake-off and the knock-off diagrams, but it does not
contain the so-called butterfly diagram (Fig.~1c in
\citet{Amusia1975jpb}), which has a vertex where the photon couples to
both electrons simultaneously. As the CCC reproduces the observed low
energy ions, it is clear that the QFM at these non-relativistic photon
energies is not dominated by the butterfly diagram, as proposed when
the QFM was first predicted \cite{Amusia1975jpb}. This is in line with a more
conventional interpretation, that refers to QFM  as a specific kinematic
region of equal energy sharing and low ion recoil, dominated by
non-dipole PDI \cite{Drukarev1995pra, Amusia2011arxiv}.
The butterfly diagram is the extreme case of SO
and TS1, when the time between the electron-electron vertex and the
photo-absorption vertex is exactly zero, which is obviously not yet
the case at 800~eV.

In conclusion, we performed kinematically complete experiments on
single photon double ionization of helium at 440 and 800~eV with
linearly polarized light. In contrast to all previous experimental work, we have
observed, for the first time, the ions with close to zero momentum, originating
from a back-to-back emission of two electrons with equal
energy sharing. This observation confirms the quasi-free mechanism
predicted nearly four decades ago by \citet{Amusia1975jpb}. Our CCC
and TDCC calculations show that these slow ions are produced by the
quadrupole interaction with the photon field and can be reproduced
without including the butterfly diagram. The newly observed double ionization
mechanism probes the two electron wave function at the
electron-electron cusp, a region previously inaccessible. These cusp
electrons can be thought of as a bosonic electron pair, which is
virtually free as they compensate completely each other's momentum, while the nucleus is at rest.
When this pair is photoionized, a total energy of 720~eV, stored in
the form of a Coulomb repulsion, is released

\begin{acknowledgments}

We thank the staff of the Advanced Light Source, in particular
H.~Bluhm and T. Tyliszczak from beamline 11.0.2.1, for their
outstanding support. M. S. Sch\"{o}ffler thanks the Alexander von
Humboldt foundation for financial support. This work is supported by the Deutsche
Forschungsgemeinschaft, DAAD, and the Office of Basic Energy Sciences,
Division of Chemical Sciences, U.S. Department of Energy, and
DOE-EPSCoR under Contracts No. DE-AC02-05CH11231 and
No. DE-FG02-07ER46357. Resources of the
Australian National Computational Infrastructure Facility were used in
this work. We thank Miron Y. Amusia for encouraging us
for 16 years to perform the present experiment.

\end{acknowledgments}

\bibliographystyle{apsrev}

\end{document}